# Approximate Dirac solutions of complex $PT$-symmetric Pöschl-Teller potential in view of spin and pseudospin symmetries

Sameer M. Ikhdair[1*], Majid Hamzavi[2**]

[1]*Physics Department, Near East University, Nicosia 922022, North Cyprus, Mersin 10, Turkey*

[2]*Department of Basic Sciences, Shahrood Branch, Islamic Azad University, Shahrood, Iran*

[*] Corresponding author *sikhdair@neu.edu.tr*
*Tel: +90-392-2236624; Fax: +90-392-2236622*
[**] *majid.hamzavi@gmail.com*

**Abstract**

By employing an exponential-type approximation scheme to replace the centrifugal term, we have approximately solved the Dirac equation for spin-$1/2$ particle subject to the complex $PT$-symmetric scalar and vector Pöschl-Teller (PT) potentials with arbitrary spin-orbit $\kappa$-wave states in view of spin and pseudospin (p-spin) symmetries. The real bound-state energy eigenvalue equation and the corresponding two-spinor components wave function expressible in terms of the hypergeometric functions are obtained by means of the wave function analysis. The spin-$1/2$ Dirac equation and the spin-$0$ Klein-Gordon (KG) equation with the complex Pöschl-Teller potentials share the same energy spectrum under the choice of $S(r) = \pm V(r)$ (i.e., exact spin and p-spin symmetries).

## 1- Introduction

A large number of potentials with real and complex forms have been studied in various fields of physics. A consistent theory of quantum mechanics (QM) in terms of Hermitian Hamiltonians is built on a complex Hamiltonian that is non-Hermitian, but the energy spectra are real as a consequence of $PT$-symmetry (parity-time reflection symmetry). The Hamiltonian is said to be $PT$-symmetric when $[PT,H]=0$ where $P$ and $T$ are the operators of parity (space) and time-reversal (complex conjugation) transformations. For a given potential $V(x)$, when we make the following transformations: $P:x\to -x$ (or $x\to a-x$), and $T:i\to -i$, $P:p\to -p$, $T:iI\to -iI$, and $PT:p\to p$, if the potential $V(-x)\to V^*(x)$ or $V(a-x)\to V^*(x)$, then the Hamiltonian is said to be $PT$-symmetric, where $x, p$ and $I$ are the position, momentum and identity operators acting in the Hilbert space. The $PT$-symmetry does not always lead to completely real spectrum and there are some cases in which part or all energy levels are found complex (see [1] and references therein).

Several potentials with complex forms have been studied in the context of the $PT$-symmetric QM by Bender *et al.* and others [2,3]. The main reason of such interest is because of the reality ($PT$-symmetry is exact) or complexity ($PT$-symmetry is spontaneously broken) of the energy spectrum of Hermitian or non-Hermitian Hamiltonians [4,5]. Recently, the $PT$-symmetry of the relativistic QM and quantum field theories [6-8] has also been investigated. The $PT$-symmetry is not a sufficient, not necessary condition for the reality of energy spectra. It has been shown by several non-hermitian $PT$-symmetric potentials like complexified Pöschl-Teller potential model (see [9] and the references therein).

Recently, the complex potentials are very popular and powerful tools of quantum mechanical calculations. In most introductory courses on quantum mechanics (QM), traditionally one is taught that the Hamiltonian operator must be Hermitian in order that the energy levels are real to describe a closed system [10], Recently, there is growing interest in the use of non-Hermitian Hamiltonians with complex potentials [10] with real energy spectra. These complex potentials extend QM to open systems where the overall probability decreases in time allowing simulations of decay, transport and scattering phenomena. Examples include modeling dissipative processes [11], nuclear [12] and chemical [13] reactions, simulation measurements of arrival, traversal or dwell times of quantum particles [14], atomic multi-photon ionization [15], electron [16] and spin transport [17]. The motivation behind the study of complex potentials is mentioned in [18]. For example, one of the motivations behind using exactly solvable quantum mechanical models (complex) $PT$-symmetric Scarf II model was to study the exactly solvable classical analogue [19]. The complex square-well potential is used by Feshbach *et al.* in the optical model of nucleus [20].

Complexification [21] is found to work successfully for harmonic oscillator potential in the second quantized version of field theory through creation and annihilation operators. Rao *et al*. [22] used complexification in their studies of ion-acoustic waves in plasma and by Yang [23] in developing a complex mechanics. The applicability and validity of the QM with complex potentials are mentioned in [24].

In this work, we will study the *PT*-symmetry of one of these potentials, the so-called Pöschl-Teller (PT) potential [25] taking the form:

$$V_{PT}(r) = \frac{\alpha^2}{2M}\left[\frac{B(B-\alpha)}{\sinh^2 \alpha r} - \frac{A(A+\alpha)}{\cosh^2 \alpha r}\right], \qquad (1)$$

where the parameter $\alpha$ is relevant to the range of the potential and $A > \alpha$, $B > +\alpha$. The PT potential is unchanged under the transformations of $A \to -(A+\alpha)$ and $B \to -(B-\alpha)$, we may only discuss the case of $B < A$. The PT potential has been extensively studied in literature [26]. It is used as the electron-nucleus potential to study the strong field ionization dynamics of a simplified one-dimensional configuration model of a homogenous molecular ion [27].

Making an imaginary coordinate shift [28], $r \to x - ix_0$, where $x_0$ is a real parameter, we obtain a non-Hermitian complex *PT*-symmetric form of the real PT potential (1) as

$$V_{CPT}(r) = \frac{\alpha^2}{2M}\left[\frac{B(B-\alpha)}{\sinh^2 \alpha(x-ix_0)} - \frac{A(A+\alpha)}{\cosh^2 \alpha(x-ix_0)}\right], \qquad (2)$$

where $x \in (0,\infty)$ and $0 < \alpha x_0 < \pi/2$. The potential (2) is a special case of the five parameter exponential-type potential model [29]. Some authors [28,29] have studied the *PT*-symmetric PT potential in the context of $s$-wave Schrödinger equation. However, the relativistic treatment of such form has not yet been reported in literature. In [30], the authors investigated the bound-states solutions of the $s$-wave Dirac equation with equally mixed PT potentials in terms of the supersymmetric quantum mechanics approach and wave function analysis. Using the same method, the p-spin symmetry solutions of the Dirac equation with PT potential for the spin-orbit quantum number $\kappa = 1$ [31] and for arbitrary spin-orbit quantum number $\kappa$ [32] have been investigated. One of us studied the exact solution of the one-dimensional KG equation for the *PT*-symmetric generalized Woods-Saxon (WS) potential using Nikiforov-Uvarov method [33]. The reality of positive and negative exact bound states of the s-states are also investigated for different types of complex generalized WS potentials [33].

The purpose of the present paper is to obtain the approximate bound-state solution of the Dirac equation with complexified $PT$-symmetric PT potential for arbitrary spin-orbit quantum number $\kappa$ under spin and p-spin symmetries by employing the wave function analysis. We shall show that if the Hamiltonian (Hermitian or non-Hermitian) has unbroken $PT$-symmetry then the energy spectrum is real. To deal with the centrifugal term $\propto r^{-2}$, we shall use exponential-type approximation form and employ an imaginary coordinate shift. To this end, we shall first briefly introduce in Section 2 the Dirac equation with radial scalar and vector potentials for arbitrary spin-orbit quantum number $\kappa$ in view of spin and p-spin symmetries. Next, the general Dirac formulas are then applied to deal with complex $PT$-symmetric scalar and vector Pöschl-Teller potentials in order to obtain the energy eigenvalues and corresponding two-component wave functions under the choice of spin and p-spin symmetries alike. Section 3 is devoted for discussions where we shall consider some particular cases for our solutions like the spin-$0$ KG and spinless Schrödinger wave equations. Finally, Section 4 contains summary and concluding remarks.

## 2. Bound State solutions

The Dirac equation for a spin-$1/2$ particle with mass $M$ moving in the field of an attractive scalar potential $S(r)$ and a repulsive vector potential $V(r)$ reads (in relativistic units $\hbar = c = 1$)

$$\left[\vec{\alpha} \bullet \vec{p} + \beta\left(M + S(r)\right)\right]\psi(\vec{r}) = \left[E - V(r)\right]\psi(\vec{r}), \tag{3}$$

where $E$ is the relativistic energy of the system, $\vec{p} = -i\vec{\nabla}$ is the three-dimensional (3D) momentum operator and $\vec{\alpha}$ and $\beta$ are the $4 \times 4$ usual Dirac matrices [34]. One may closely follow the procedure described in Eqs. (13)-(19) of Ref. [34] to obtain

$$\left\{\frac{d^2}{dr^2} - \frac{\kappa(\kappa+1)}{r^2} - \left[M + E_{n\kappa} - \Delta(r)\right]\left[M - E_{n\kappa} + \Sigma(r)\right]\right.$$

$$\left. + \frac{\frac{d\Delta(r)}{dr}}{M + E_{n\kappa} - \Delta(r)}\left(\frac{d}{dr} + \frac{\kappa}{r}\right)\right\} F_{n\kappa}(r) = 0, \quad \kappa(\kappa+1) = l(l+1),\ r \in (0, \infty) \tag{4}$$

$$\left\{\frac{d^2}{dr^2}-\frac{\kappa(\kappa-1)}{r^2}-\left[\left(M+E_{n\kappa}-\Delta(r)\right)\left(M-E_{n\kappa}+\Sigma(r)\right)\right]\right.$$

$$\left.-\frac{\dfrac{d\Sigma(r)}{dr}}{M-E_{n\kappa}+\Sigma(r)}\left(\frac{d}{dr}-\frac{\kappa}{r}\right)\right\}G_{n\kappa}(r)=0,\ \kappa(\kappa-1)=\tilde{l}\left(\tilde{l}+1\right),\ r\in(0,\infty) \qquad (5)$$

where $\Delta(r)=V(r)-S(r)$ and $\Sigma(r)=V(r)+S(r)$. The orbit-spin quantum number $\kappa$ is related to the orbital quantum numbers $l$ and $\tilde{l}$ for spin and p-spin symmetric models, respectively, as

$$\kappa=\begin{cases}-(l+1)=-(j+\dfrac{1}{2})\ (s_{1/2},p_{3/2},etc.)\ j=l+\dfrac{1}{2},\ \text{aligned spin}\ (\kappa<0),\\ +l=+(j+\dfrac{1}{2})\ \ \ (p_{1/2},d_{3/2},etc.)\ j=l-\dfrac{1}{2},\ \text{unaligned spin}\ (\kappa>0).\end{cases}$$

Further, $\kappa$ in the quasi-degenerate doublet structure can be expressed in terms of $\tilde{s}=1/2$ and $\tilde{l}$, the p-spin and pseudo-orbital angular momentum, respectively, as

$$\kappa=\begin{cases}-\tilde{l}=-(j+\dfrac{1}{2})\ \ \ (s_{1/2},p_{3/2},etc.)\ j=\tilde{l}-\dfrac{1}{2},\ \text{aligned pseudospin}\ (\kappa<0),\\ +(\tilde{l}+1)=+(j+\dfrac{1}{2})\ (d_{3/2},f_{5/2},etc.)\ j=\tilde{l}+\dfrac{1}{2},\ \text{unaligned spin}\ (\kappa>0),\end{cases}$$

where $\kappa=\pm 1,\pm 2,\ldots$. For example, the states $\left(1s_{1/2},0d_{3/2}\right)$ and $\left(1p_{3/2},0f_{5/2}\right)$ can be considered as p-spin doublets.

## 2.1. Spin Symmetric Limit

Let us consider the exact spin symmetry, $\dfrac{d\Delta(r)}{dr}=0$ or $\Delta(r)=C_s=\text{constant}$ [35,36]. In Eq. (4), we set the sum potential $\Sigma(r)$ as the $PT$-symmetric Pöschl-Teller potential, i.e.,

$$\Sigma(r)=2V_{\text{CPT}}(r)=\frac{\alpha^2}{M}\left[\frac{B(B-\alpha)}{\sinh^2\alpha r}-\frac{A(A+\alpha)}{\cosh^2\alpha r}\right],$$

and setting $r\to x-ix_0$; where $x\in(0,\infty)$ is real part and $x_0$ is a constant real number, we obtain

$$\left\{\frac{d^2}{dx^2}+\beta^2-\frac{\kappa(\kappa+1)}{r^2}-\frac{\alpha^2}{M}\left[\frac{B(B-\alpha)}{\sinh^2\alpha(x-ix_0)}-\frac{A(A+\alpha)}{\cosh^2\alpha(x-ix_0)}\right]M_s\right\}F_{n\kappa}(x)=0, \quad (6a)$$

$$M_s = M + E_{n\kappa} - C_s \text{ and } \beta^2 = E_{n\kappa}^2 - M^2 + C_s(M - E_{n\kappa}), \quad (6b)$$

where $\kappa = l$ and $\kappa = -l-1$ for $\kappa < 0$ and $\kappa > 0$, respectively. Further $F_{n\kappa}(r) \equiv F_{n,\kappa}(x)$ is used. Note that the above second order differential Schrödinger-type equation can be solved exactly only for $s$-wave $(\kappa = -1)$ case This equation cannot be solved analytically for $\kappa \neq 0$ due to the centrifugal term $\kappa(\kappa+1)r^{-2}$. In order to obtain approximate analytical solution of Eq. (6a) for the case of non-zero $\kappa$ values, we use an approximation for the centrifugal term similar to the one used in [37,38]. We take the following approximation for the centrifugal term [34,37,38]

$$\frac{\kappa(\kappa+1)}{r^2} = \kappa(\kappa+1)\lim_{\alpha\to 0}4\alpha^2\left[d_0+\frac{1}{\left(e^{\alpha r}-e^{-\alpha r}\right)^2}\right] \approx \alpha^2\kappa(\kappa+1)\left[4d_0+\frac{1}{\sinh^2\alpha r}\right], \quad (7)$$

where $d_0 = 1/12$ is a dimensionless constant. Such an approximation is a good approximation for small values of screening parameter $\alpha$, i.e., $\alpha r \ll 1$. Note that this approximation turns to become the conventional approximation introduced by Greene and Aldrich [39] and used by one of us [40] when $d_0 = 0$. In the present relativistic case, the treatment of this term does not raise any problem as in the nonrelativistic case [37] since the centrifugal term (7) can be reduced to $l(l+1)/r^2$ for $\kappa < 0$ and $\kappa > 0$ and hence the non-relativistic techniques are still valid here. Therefore, inserting this approximation into Eq. (6a) and applying the change of variables $r \to x - ix_0$, we can recast the Schrödinger-type equation for the upper spinor component as

$$\left\{\frac{d^2}{dx^2}+\tilde{E}_{n\kappa}^2-\frac{V_2}{\sinh^2\alpha(x-ix_0)}+\frac{V_1}{\cosh^2\alpha(x-ix_0)}\right\}F_{n\kappa}(x)=0, \quad (8)$$

with the identifications:

$$V_1 = \frac{\alpha^2 A(A+\alpha)M_s}{M}, \quad (9a)$$

$$V_2 = \alpha^2\left(\kappa(\kappa+1)+\frac{B(B-\alpha)M_s}{M}\right), \quad (9b)$$

$$\tilde{E}_{n\kappa}^2 = \beta^2 - 4\alpha^2\kappa(\kappa+1)d_0. \quad (9c)$$

The deformed hyperbolic functions [41,42] are defined by

$$\sinh_{q_c}(\alpha x) = \frac{e^{\alpha x} - q_c e^{-\alpha x}}{2}, \quad \cosh_{q_c}(\alpha x) = \frac{e^{\alpha x} + q_c e^{-\alpha x}}{2}, \quad q_c = e^{i2\alpha x_0} \tag{10a}$$

$$\sinh_{q_c}(2\alpha x) = \frac{e^{2\alpha x} - q_c e^{-2\alpha x}}{2}, \quad \cosh_{q_c}(2\alpha x) = \frac{e^{2\alpha x} + q_c e^{-2x}}{2}, \quad (x \to 2x) \tag{10b}$$

where $q_c$ is a complex number. Now, we try to explore real solution for Eq. (8). By using the above definitions and further making a little algebra, one can easily show the relations:

$$\sinh \alpha(x - ix_0) = \frac{1}{\sqrt{q_c}} \sinh_{q_c}(\alpha x), \quad \cosh \alpha(x - ix_0) = \frac{1}{\sqrt{q_c}} \cosh_{q_c}(\alpha x), \tag{11a}$$

$$2\cosh^2_{q_c}(\alpha x) = q_c + \sinh_q(2\alpha x), \tag{11b}$$

$$\cosh^2_{q_c}(\alpha x) - \sinh^2_{q_c}(\alpha x) = q_c, \tag{11c}$$

$$\cosh^2_{q_c}(\alpha x) + \sinh^2_{q_c}(\alpha x) = \sinh_q(2\alpha x), \tag{11d}$$

$$4\sinh^2_{q_c}(\alpha x)\cosh^2_{q_c}(\alpha x) = \cosh^2_q(2\alpha x), \tag{11e}$$

$$\cosh^2_{q_c}(2\alpha x) - \sinh^2_{q_c}(2\alpha x) = q_c, \quad (x \to 2x) \tag{11f}$$

where $q = -q_c^2$. Inserting the relations in Eq. (11) into Eq. (8), we can obtain

$$\left\{ \frac{d^2}{dx^2} + \tilde{E}^2_{n\kappa} - U_{\text{eff}}(x) \right\} F_{n\kappa}(x) = 0, \tag{12a}$$

$$U_{\text{eff}}(x) = \frac{1}{\cosh^2_q(2\alpha x)} \left[ 2q_c^2(V_1 + V_2) + 2q_c(V_2 - V_1)\sinh_q(2\alpha x) \right], \tag{12b}$$

where $U_{\text{eff}}(x)$ is the effective $PT$-symmetric Pöschl-Teller potential. Now, we need to solve Eq. (12). This can be done by defining a new variable

$$2z = 1 - \frac{i}{\sqrt{q}} \sinh_q(2\alpha x), \quad \frac{d}{dx} = -\frac{i\alpha}{\sqrt{q}} \cosh_q(2\alpha x) \frac{d}{dz},$$

$$\frac{d^2}{dx^2} = -\frac{\alpha^2}{q} \left( \cosh^2_q(2\alpha x) \frac{d^2}{dz^2} + 2i\sqrt{q} \sinh_q(2\alpha x) \frac{d}{dz} \right), \tag{13}$$

where $q_c = i\sqrt{q}$. Furthermore, making use of Eq. (10b), we can obtain $\cosh^2_q(2\alpha x) - \sinh^2_q(2\alpha x) = q$. Using Eq. (13), we can rearrange Eq. (12) in a more simple form:

$$z(1-z) \frac{d^2 F_{n\kappa}(z)}{dz^2} + \left( \frac{1}{2} - z \right) \frac{dF_{n\kappa}(z)}{dz} - \frac{1}{4\alpha^2} \left[ \tilde{E}^2_{n\kappa} + \frac{V_1}{z(1-z)} + \frac{(V_2 - V_1)}{(1-z)} \right] F_{n\kappa}(z) = 0, \tag{14}$$

where $z \in (-\infty, 0)$. Further, inserting the appropriate ansatz of the wave function

$$F_{n\kappa}(z) = z^{-\lambda}(1-z)^{-\eta} f_{n\kappa}(z), \tag{15}$$

into Eq. (14), leads to the following hypergeometric differential equation [43],

$$z(1-z)\frac{d^2 f_{n\kappa}(z)}{dz^2} + \left[\frac{1}{2} - 2\lambda - (-2\lambda - 2\eta + 1)z\right]\frac{df_{n\kappa}(z)}{dz}$$
$$-\left[(\lambda+\eta)^2 + \left(\frac{\tilde{E}_{n\kappa}}{2\alpha}\right)^2 + \frac{c_1}{z(1-z)} + \frac{c_2}{(1-z)}\right] f_{n\kappa}(z) = 0, \tag{16}$$

where

$$c_1 = -\lambda^2 - \frac{\lambda}{2} + \frac{V_1}{4\alpha^2}, \quad c_2 = -\eta^2 - \frac{\eta}{2} + \frac{V_2}{4\alpha^2}. \tag{17}$$

Equation (16) can be reduced to a hypergeometric equation when taking $c_1 = c_2 = 0$. In solving (17), we obtain

$$\lambda = \frac{1}{4}\left(-1 + \sigma\sqrt{1 + \frac{4V_1}{\alpha^2}}\right), \tag{18a}$$

$$\eta = \frac{1}{4}\left(-1 + \tau\sqrt{1 + \frac{4V_2}{\alpha^2}}\right), \tag{18b}$$

where $\sigma = \pm 1$ and $\tau = \mp 1$. Hence, the solution of this equation is simply the hypergeometric function:

$$f_{n\kappa}(x) = {}_2F_1(a,b;c;z) = \sum_{\gamma=0}^{\infty} \frac{(a)_\gamma (b)_\gamma}{(c)_\gamma} \frac{x^\gamma}{\gamma!}$$
$$= {}_2F_1\left(-\lambda - \eta + i\frac{\tilde{E}_{n\kappa}}{2\alpha}, -\lambda - \eta - i\frac{\tilde{E}_{n\kappa}}{2\alpha}; \frac{1}{2} - 2\lambda; 1 - iq^{-1/2}\sinh_q(2\alpha x)\right), \tag{19}$$

where $(a)_m = \Gamma(a+m)/\Gamma(a)$ is a Pochhammer symbol. When either $a$ or $b$ equals a negative integer $-n$, the hypergeometric function $f_{n\kappa}(x)$ can be reduced to a polynomial of degree $n$ and asymptotically vanishing under certain boundary conditions. This shows that the hypergeometric function given in Eq. (19) can be finite under the following quantum condition:

$$-\lambda - \eta + i\frac{\tilde{E}_{n\kappa}}{2\alpha} = -n, \quad n = 0, 1, 2, \cdots, \tag{20}$$

from which, together with Eqs. (6b), (9c) and (18), we finally obtain the energy eigenvalue equation for the nuclei in the field of relativistic $PT$-symmetric Pöschl-Teller potential in view of the spin symmetry,

$$M^2 - E_{n\kappa}^2 + C_s(E_{n\kappa} - M) = -\frac{\alpha^2 \kappa(\kappa+1)}{3} + 4\alpha^2$$

$$\times \left[ -n - \frac{1}{2} + \frac{1}{4}\left( \sigma\sqrt{1 + \frac{4(M + E_{n\kappa} - C_s)A(A+\alpha)}{M}} + \tau\sqrt{(2\kappa+1)^2 + \frac{4(M + E_{n\kappa} - C_s)B(B-\alpha)}{M}} \right) \right]^2, \quad (21)$$

where $A > B$. Note that the following two choices of $\sigma = -1$ and $\tau = +1$ or $\sigma = 1$ and $\tau = -1$ give same numerical results with $\lambda > 0$, $\eta < 0$, $\lambda < -\eta$ and $\lambda, \eta \in \mathbb{R}$. The energy levels $E_{n\kappa}$ are defined implicitly by energy eigenvalue equation (21) which is a rather complicated transcendental energy equation. From Eq. (21), it can be seen that the complexified $PT$-symmetric PT (2) possesses real energy eigenvalue equation (21) as the real PT potential (1) with spin symmetric choice. Further, we can find the maximum (largest) level $n_{\max}$ from the reality condition of the energy spectrum (21) as the largest integer that is less than or equal to [44]

$$n_{\max} = \frac{1}{4}\left( \pm\sqrt{1 + \frac{4(M + E_{n\kappa} - C_s)A(A+\alpha)}{M}} \mp \sqrt{(2\kappa+1)^2 + \frac{4(M + E_{n\kappa} - C_s)B(B-\alpha)}{M}} \right) - \frac{1}{2},$$

or equivalently the reality of (21) provides $E_{n\kappa} = E_{n_{\max}\kappa}$. Hence, Eq. (21) satisfies the condition, $0 < n < n_{\max} = \lambda + \eta$, $n = 0, 1, 2, \cdots, \lambda + \eta$.

In addition, note that energy spectrum formula (21) looks same as Eq. (24) of Ref. [45] in KG equation case with $S(r) = V(r)$ (exact spin symmetry, $C_s = 0$) real Pöschl-Teller potentials after making changes for the PT potential parameters in Eq. (1), i.e., $A(A+\alpha) \to \lambda(\lambda+1)$ and $B(B-\alpha) \to k(k-1)$. This means that the two forms (real and complex $PT$-symmetry) of the PT potential possess the same real energy spectrum under spin symmetry.

The corresponding upper-spinor component $F_{n\kappa}(x)$ is expressible in the form of the hypergeometric function as

$$F_{n\kappa}(x) = N_{n\kappa} 2^{\lambda+\eta} (p_1(x))^{-\lambda} (p_2(x))^{-\eta} {}_2F_1\left(-n, -2(\lambda+\eta)+n; -2\lambda+\frac{1}{2}; \frac{p_1(x)}{2}\right), \quad (22)$$

where $p_1(x) = 1 - e^{-i2\alpha x_0}\sinh_q(2\alpha x)$, $p_2(x) = 1 + e^{-i2\alpha x_0}\sinh_q(2\alpha x)$ and $N_{n\kappa}$ is the normalization constant. The wave function should vanish under certain asymptotic behavior of the wave function, i.e., $F_{n\kappa}(x \to \pm\infty) = 0$. We can use the following recurrence relation between hypergeometric functions:

$$\frac{d}{dz}\left({}_2F_1(a,b;c;z)\right) = \left(\frac{ab}{c}\right){}_2F_1(a+1,b+1;c+1;z), \tag{23}$$

and express the hypergemetric function in terms of the Jacobi polynomials:

$${}_2F_1\left(-n,-2(\lambda+\eta)+n;-2\lambda+\frac{1}{2};\frac{p_1(x)}{2}\right)$$

$$= \frac{(-2\lambda+1/2)_n}{n!} P_n^{(-2\lambda-1/2,-2\eta-1/2)}\left(e^{-i2\alpha x_0}\sinh_q(2\alpha x)\right),$$

in finding the lower component, $G_{n\kappa}(x)$ which can be obtained from Eq. (18) of Ref. [34] by using Eq. (22) as

$$G_{n\kappa}(x) = \frac{1}{(M+E_{n\kappa}-C_s)}\left[\frac{\kappa}{r}F_{n\kappa}(x)+\frac{dF_{n\kappa}(x)}{dr}\right]$$

$$= \frac{2n(n-2\eta-2\lambda)}{(1-4\lambda)(M+E_{n\kappa}-C_s)}\alpha e^{-i2\alpha x_0}\cosh_q(2\alpha x)$$

$$\times {}_2F_1\left(1-n,1+n-2(\lambda+\eta);-2\lambda+\frac{3}{2};\frac{p_1(x)}{2}\right)(p_1(x))^{-\lambda}(p_2(x))^{-\eta}$$

$$+\frac{F_{n\kappa}(x)}{(M+E_{n\kappa}-C_s)}\left[\frac{\kappa}{x-ix_0}+2\alpha e^{-i2\alpha x_0}\cosh_q(2\alpha x)\left(\frac{\lambda}{p_1(x)}-\frac{\eta}{p_2(x)}\right)\right], \lambda \neq \frac{1}{4}. \tag{24}$$

From Eq. (24), we know that in the limit of exact spin symmetry there are only bound positive energy states, otherwise the lower spinor component $G_{n\kappa}(x)$ in (24) will diverge if $E_{n\kappa}=-M$ and $C_s=0$. There are no bound negative energy states in view of spin symmetry condition.

The two-spinors $F_{n\kappa}(x)$ and $G_{n\kappa}(x)$ satisfy the regularity boundary conditions for the bound states when $\lambda > 0$, $\eta < 0$, and $\lambda, \eta \in \mathbb{R}$. Considering the limiting case $\alpha \to 0$, we find from Eq. (21) that the energy eigenvalue approaches a constant value, i.e., $\lim_{\alpha\to 0} E_{n\kappa} = M$ or $\lim_{\alpha\to 0} E_{n\kappa} = C_s - M$. However, the energy limit $\lim_{\alpha\to 0} E_{n\kappa} = C_s - M$. is not physically acceptable since it makes the lower component $G_{n\kappa}(x)$ diverge. Therefore, we choose the energy limit $\lim_{\alpha\to 0} E_{n\kappa} = M$ as the physically acceptable one and obtain

$$\lim_{\alpha\to 0} G_{n\kappa}(x) = \left(\frac{\kappa}{x-ix_0}\right)\frac{\lim_{\alpha\to 0} F_{n\kappa}(x)}{(2M-C_s)}.$$

To show the procedure of determining the bound-state energy eigenvalues from Eq. (21), we take a set of physical parameter values, $\alpha = 0.35\, fm^{-1}$, $A=8$, $B=2$, $M = 5.0 \, fm^{-1}$ and $C_s = -0.35 \, fm^{-1}$, to give a numerical example. When $n = 0$ and $\kappa = 1$ or $-2$, Eq. (21) yields two values for energy, however, we select the positive energy as the physical solution for the finiteness of the lower component of the wave function (24) and (22) for the upper component of the wave function. If we take $E_{0,1} = 4.320628792 \, fm^{-1}$ as the solution of Eq. (21), and find the values of $\lambda$ and $\eta$ are $\lambda = 4.989398388$ and $\eta = -1.634030092$, which calculate the value $n_{max} = 3.550238160$. With the same parameter values $\alpha$, $A$, $B$, $M$ and $C_s$, the numerical solutions of Eq. (21) for the other values of $n$ and $\kappa$ are presented in Table 1. This Table shows the spin partners, i.e., the Dirac eigenstates $0p_{3/2}$ and $0p_{1/2}$. We have used $\sigma = 1$, $\tau = -1$ or $\sigma = -1$, $\tau = 1$ in calculating spin symmetric spectrum. Further, in Figure 1 to Figure 5, we plot the variation of the spin symmetric energy eigenvalues as function of the parameters $\alpha$, $A$, $B$, $M$ and $C_s$. We find out that $E_{n\kappa}$ becomes more positive (increases) as the parameter values of $M, B$ and $C_s$ increase. However, it becomes less positive (decreases) as the parameter values of $\alpha$ and $A$ decrease.

## 2.2. P-spin Symmetric Limit

The p-spin symmetry occurs when the relationship between the vector potential and the scalar potential is given by $V(r) = -S(r)$ [46]. Further, if $\dfrac{d[V(r) + S(r)]}{dr} = \dfrac{d\Sigma(r)}{dr} = 0$, then $\Sigma(r) = C_{ps} =$ constant, for which the p-spin symmetry is exact in the Dirac equation [47,48]. Thus, taking the potential difference $\Delta(r)$ as the $PT$-symmetric potential, i.e.,

$$\Delta(r) = 2V_{\text{CPT}}(r) = \frac{\alpha^2}{M}\left[\frac{B(B-\alpha)}{\sinh^2 \alpha r} - \frac{A(A+\alpha)}{\cosh^2 \alpha r}\right],$$

and setting $r = x - ix_0$; Eq. (5) under this symmetry becomes

$$\left\{\frac{d^2}{dx^2} + \tilde{\bar{E}}_{n\kappa}^2 - \frac{\tilde{V}_2}{\sinh^2 \alpha(x - ix_0)} + \frac{\tilde{V}_1}{\cosh^2 \alpha(x - ix_0)}\right\} G_{n\kappa}(x) = 0,$$

$M_{ps} = M - E_{n\kappa} + C_{ps}$ and $\tilde{\beta}^2 = E_{n\kappa}^2 - M^2 - C_{ps}(M + E_{n\kappa})$, (25)

where

$$\tilde{V}_1 = \frac{\alpha^2 A(A+\alpha)M_{ps}}{M}, \tag{26a}$$

$$\tilde{V}_2 = \alpha^2 \left( \kappa(\kappa-1) + \frac{B(B-\alpha)M_{ps}}{M} \right), \tag{26b}$$

$$\bar{\tilde{E}}_{n\kappa}^2 = \tilde{\beta}^2 - 4\alpha^2 \kappa(\kappa-1)d_0. \tag{26c}$$

To avoid repetition, the negative energy solution of Eq. (5), in the p-spin symmetric limit: $V(r) = -S(r)$, can be easily obtained directly via the spin symmetric solution through the parametric mappings [49-53]:

$$F_{n\kappa}(x) \leftrightarrow G_{n\kappa}(x),\ \kappa \to \kappa-1,\ V(x) \to -V(x),\ E_{n\kappa} \to -E_{n\kappa},\ C_s \to -C_{ps}. \tag{27}$$

In following the previous procedure, one can easily obtain the energy eigenvalue equation for the nuclei in the field of the relativistic complex $PT$-symmetric Pöschl-Teller potential (2) under the exact p-spin symmetry limit,

$$M^2 - E_{n\kappa}^2 + C_{ps}\left(E_{n\kappa} + M\right) = -\frac{\alpha^2 \kappa(\kappa-1)}{3} + 4\alpha^2$$

$$\times \left[ -n - \frac{1}{2} + \frac{1}{4}\left( \sigma\sqrt{1 - \frac{4(M - E_{n\kappa} + C_{ps})A(A+\alpha)}{M}} + \tau\sqrt{(2\kappa-1)^2 - \frac{4(M - E_{n\kappa} + C_{ps})B(B-\alpha)}{M}} \right) \right]^2, \tag{28}$$

where the following two choices of $\sigma = +1$ and $\tau = -1$ or $\sigma = -1$ and $\tau = +1$ give same numerical results as in spin symmetric case. Hence, Eq. (28) satisfies the restriction condition:

$$n_{\max} = \frac{1}{4}\sqrt{1 - \frac{4(M - E_{n\kappa} + C_{ps})A(A+\alpha)}{M}} - \frac{1}{4}\sqrt{(2\kappa-1)^2 - \frac{4(M - E_{n\kappa} + C_{ps})B(B-\alpha)}{M}} - \frac{1}{2},$$

where $0 < n < n_{\max} = \nu + \delta,\ n = 0,\ 1,\ 2,\cdots,\ \nu + \delta.$

The energy spectrum equation (28) is identical to Eq. (10) of Ref. [32] obtained by Akçay for the real PT potential (1) using the conventional approximation scheme [39]. This means that the two forms (real and complex $PT$-symmetry) of the PT potential possess the same real energy spectrum in the exact p-spin symmetry. On the other hand, the lower-spinor component of the wave functions is found as

$$G_{n\kappa}(x) = 2^{\nu+\delta}\tilde{N}_{n\kappa}\left(p_1(x)\right)^{-\nu}\left(p_2(x)\right)^{-\delta}{}_2F_1\left(-n, -2(\nu+\delta)+n; -2\delta + \frac{1}{2}; \frac{p_1(x)}{2}\right), \tag{29}$$

where

$$\nu = -\frac{1}{4}\left(1 - \sqrt{1 - \frac{4(M - E_{n\kappa} + C_{ps})A(A+\alpha)}{M}}\right), \tag{30a}$$

$$\delta = -\frac{1}{4}\left(1 + \sqrt{(2\kappa-1)^2 - \frac{4(M - E_{n\kappa} + C_{ps})B(B-\alpha)}{M}}\right), \tag{30b}$$

where $\tilde{N}_{n\kappa}$ is the normalization constant. The upper-spinor component of the Dirac wave function can be calculated by

$$F_{n\kappa}(r) = \frac{1}{M - E_{n\kappa} + C_{ps}}\left(\frac{d}{dr} - \frac{\kappa}{r}\right)G_{n\kappa}(r), \tag{31}$$

where $E \neq M + C_{ps}$ and with exact p-spin symmetry ($C_{ps} = 0 \to S(r) = -V(r)$), only negative energy states do exist. The upper component $F_{n\kappa}(r)$ can be obtained as

$$F_{n\kappa}(x) = \frac{2n(n - 2\nu - 2\delta)2^{\nu+\delta}}{(1-4\nu)(M - E_{n\kappa} + C_{ps})} \alpha e^{-i2\alpha x_0} \cosh_q(2\alpha x)$$

$$\times {}_2F_1\left(1-n, 1+n-2(\nu+\delta); -2\nu+\frac{3}{2}; \frac{1}{2}p_1(x)\right)(p_1(x))^{-\nu}(p_2(x))^{-\delta}$$

$$+ \frac{G_{n\kappa}(x)2^{\nu+\delta}}{(M - E_{n\kappa} + C_{ps})}\left[\frac{\kappa}{x - ix_0} + 2\alpha e^{-i2\alpha x_0}\cosh_q(2\alpha x)\left(\frac{\nu}{p_1(x)} - \frac{\delta}{p_2(x)}\right)\right], \quad \nu \neq \frac{1}{4}, \tag{32}$$

where $E_{n\kappa} \neq M$ when $C_{ps} = 0$. Note that the singularity of the upper-component $F_{n\kappa}(x)$ at $E_{n\kappa} \neq M$ demands us to choose only the negative energy solution for the sake of the normalizability of the two-spinor components of the p-spin wave function. To generate the binding energy spectrum of Eq. (28), we take a set of parameter values, $\alpha = 0.35\,fm^{-1}$, $A = 8$, $B = 2$, $M = 5.0\,fm^{-1}$ and $C_{ps} = -10.0\,fm^{-1}$. With the choice of negative energy eigenvalues as physical solution to Eq. (32), we present the energy spectrum for the p-spin case in Table 2. Further, in Figure 6 to Figure 10, we plot the variation of the p-spin symmetric energy eigenvalues as function of the parameters $\alpha$, $A$, $B$, $M$ and $C_{ps}$. We find out that $E_{n\kappa}$ becomes more negative (more attractive) as the parameter values of $M, A$ and $\alpha$ increase. However, it becomes less negative (less attractive) as the parameter values of $C_{ps}$ and $B$ decrease.

## 3. Discussions

When the $PT$-symmetric Pöschl-Teller potential is taken in the form [45]

$$V_{PT}(r) = \frac{\alpha^2}{2M}\left[\frac{k(k-1)}{\sinh^2\alpha(x - ix_0)} - \frac{\lambda(\lambda+1)}{\cosh^2\alpha(x - ix_0)}\right], \quad k > 1,\ \lambda > 1, \tag{33}$$

then it remains unchanged under the transformations of $\lambda \to -\lambda -1$ and $k \to -k+1$.
The potential (33) has the approximate energy spectrum formula in arbitrary $\kappa$-state, (in relativistic $\hbar = c = 1$ units)

$$M^2 - E_{n\kappa}^2 + C_s(E_{n\kappa} - M) = -\frac{\alpha^2 \kappa(\kappa+1)}{3} + 4\alpha^2$$

$$\times \left[ -n - \frac{1}{2} \pm \frac{1}{4}\sqrt{1 + \frac{4(M + E_{n\kappa} - C_s)\lambda(\lambda+1)}{M}} \mp \frac{1}{4}\sqrt{(2\kappa+1)^2 + \frac{4(M + E_{n\kappa} - C_s)k(k-1)}{M}} \right]^2,$$

(34)

which is identical to Eq. (42) of Ref. [45] found for real PT potential. Making the replacements of $C_s = 0$ and $\kappa(\kappa+1) = l(l+1)$, one can readily find out that the Dirac equation and KG equation share the same energy spectrum under the choice of equally mixed radial scalar and vector potentials, i.e., $S(r) = V(r)$ for the $PT$-symmetric PT potentials as remarked in Ref. [54] (see, e.g., Eq. (24) of Ref. [45]). The energy equation of (33) in the Dirac equation under the p-spin symmetry,

$$M^2 - E_{n\kappa}^2 + C_{ps}(E_{n\kappa} + M) = -\frac{\alpha^2 \kappa(\kappa-1)}{3} + 4\alpha^2$$

$$\times \left[ -n - \frac{1}{2} \pm \frac{1}{4}\sqrt{1 - \frac{4(M - E_{n\kappa} + C_{ps})\lambda(\lambda+1)}{M}} \mp \frac{1}{4}\sqrt{(2\kappa-1)^2 - \frac{4(M - E_{n\kappa} + C_{ps})k(k-1)}{M}} \right]^2,$$

(35)

which is same for the solution of the real PT potential in Eq. (44) of [45]. We observe that the Dirac and KG equations share the same energy spectrum (see Eq. (36) of Ref. [45]) when $S(r) = -V(r)$, $C_{ps} = 0$ and $\kappa(\kappa-1) = l(l+1)$ [54].

In the nonrelativistic limits

$$\kappa(\kappa+1) = l(l+1),\ C_s = 0,\ E_{n\kappa} + M \to 2\mu \text{ and } E_{n\kappa} - M \to E_{nl},\qquad(36)$$

then Eq. (34) becomes

$$E_{nl} = \frac{\alpha^2 l(l+1)d_0}{2\mu} - \frac{2\alpha^2}{\mu}\left[ -n - \frac{1}{2} + \frac{1}{4}\sqrt{1 + 8\lambda(\lambda+1)} + \frac{1}{4}\sqrt{(1+2l)^2 + 8k(k-1)} \right]^2,\qquad(37)$$

where $d_0 = 1/12$.

## 4. Summary and Concluding Remarks

In this paper, we have investigated the bound state solutions of the Dirac equation with $PT$-symmetric PT potential for any spin-orbit quantum number $\kappa$. By using an approximation scheme to deal with the centrifugal term and making a complex transformation in coordinates, we have obtained the energy eigenvalue equation and the unnormalized upper- and lower-spinor components of the radial wave function expressible in terms of the Jacobi polynomials in view of spin symmetry with any $\kappa$-wave state. It is noticed that the complexified $PT$-symmetric PT potential carries the same real energy eigenvalue solutions as the real PT in the spin-$1/2$ Dirac and spin-$0$ KG equations. However, the wave functions of the complexified $PT$-symmetric PT potential have different asymptotic behavior than the ones for the real PT potential. Furthermore, in obtaining the p-spin symmetric solutions from the spin symmetric ones, we have employed parametric mappings [49-53]. The complexified $PT$-symmetric potential may have real (complex) energy spectrum when the wave function is normalizable (not normalizable) in a given range as recently stated in our recent work [55].

**Acknowledgments**

We thank the kind referees for their invaluable suggestions and critics that have greatly helped in improving this paper. We would also like to thank the editors for their patience and kind cooperation. Sameer M. Ikhdair acknowledges the partial support of the Scientific and Technological Research Council of Turkey.

**References**

[1] S. M. Ikhdair and R. Sever, Int. J. Theor. Phys. **46** (2007) 1643; S. M. Ikhdair and R. Sever, Int. J. Mod. Phys. E **17** (6) (2008) 1107.

[2] C. M. Bender and S. Boettcher, Phys. Rev. Lett. **80** (1998) 5243.

[3] C. M. Bender and S. Boettcher, J. Phys. A **31** (1998) 1273.

[4] C. M. Bender, D. C. Brody and H. F. Jones, Phys. Rev. Lett. **89** (2002) 270402.

[5] C. M. Bender, G. V. Dunne, P. N. Meisenger and M. Şimşek, Phys. Lett. A **281** (2001) 311.

[6] C. M. Bender, D. C. Brody and H. F. Jones, Phys. Rev. Lett. **93** (2004) 251601.

[7] C. M. Bender, H. F. Jones and R. J. Rivers, Phys. Lett. B **625** (2005) 333.

[8] C. M. Bender, D. C. Brody, J. H. Chen, H. F. Jones, K. A. Milton and M.C. Ogilvie, Phys. Rev. D **74** (2006) 025016; A. de Souza Dutra, V. G. C. S. dos Santos and A. C. Amaro de Faria, Jr., Phys. Rev. D **75** (2007) 125001.

[9] C. -S. Jia, Y. Sun and Y.Li, Phys. Lett. A **305** (2002) 231.

[10] C. M. Bender, Rep. Prog. Phys. **70** (2007) 947 [arXiv: hep.th/0703096]; Ingrid Rotter, J. Phys. A: Math. Theor. **42** (2009) 153001; C. E. Ruter *et al*. Nat. Phys. **6** (2010) 192; J. Dalibard, Y. Castin and K. Molmer, Phys. Rev. Lett. **68** (1992) 580.

[11] D. K. Ferry and J. R. Barker, Appl. Phys. Lett. **74** (1999) 582.

[12] P. E. Hodgson, Rep. Prog. Phys. **34** (1971) 765.

[13] W. H. Miller, Science **233** (1986) 171.

[14] A. Ruschhaupt, J. A. Damborenea, B. Navarro, J. G. Muga and G. C. Hegerfeldt, Europhys. Lett. **67** (2004) 1.

[15] H. C. Baker, Phys. Rev. A **30** (1984) 773.

[16] A. Aviram and M. A. Ratner, Chem. Phys. Lett. **29** (1974) 277.

[17] F. Doğan, W. Kim, C. M. Blois and F. Marsiglio, Phys. Rev. B **77** (2008) 195107.

[18] R. S. Kaushal, Pramana- J. Phys. **73** (2009) 287.

[19] A. Sinha, D. Dutta and P. Roy, Phys. Lett. A **375** (2011) 452; Z. Ahmed, Phys. Lett. A **290** (2001) 19.

[20] H. Feshbach, C. E. Porter and V. F. Weisskopf, Phys. Rev. **96** (1954) 488.

[21] R. K. Colegrave, P. Croxson and M. A. Marran, Phys. Lett. A **131** (1988) 407.

[22] N. N. Rao, B. Buti and S. B. Khadkikar, Pramana- J. Phys. **27** (1986) 497; B. Buti, N. N. Rao and S. B. Khadkikar, Phys. Scr. **34** (1986) 729.

[23] C. –D. Yang, Chaos, Solitons and Fractals **33** (2007) 1073.

[24] B. D. Wibking and K. Varga, Phys. Lett. A **376** (2012) 365.

[25] G. Pöschl and E. Teller, Z. Physik **83** (1933) 143.

[26] G. F. Wei and S. H. Dong, Phys. Lett. A **373** (2009) 2428.

[27] Y. S. Chooi and C. B. Moore, J. Chem. Phys. **110** (1993) 1111.

[28] M. Znojil, J. Phys. A : Math. Gen. **33** (2000) 4561.

[29] C. S. Jia, Y. Li, Y. Sun, J. Y.Liu and L. T. Sun, Phys. Lett. A **311** (2003) 115.

[30] X. C. Zhang, Q. W. Liu, C. S. Jiang and L. Z. Wang, Phys. Lett. A **340** (2005) 59.

[31] C. S. Jia, P. Guo, Y. F. Diao, L. Z. Yi and X. J. Xie, Eur. Phys. J. A **34** (2007) 41.

[32] Y. Xu, S. He and C. S. Jia, J. Phys. A: Math. Theor. **41** (2008) 255302; H. Akçay, J. Phys. A: Math. Theor. **42** (2009) 198001.

[33] S. M. Ikhdair and R. Sever, Ann. Phys. (Leipzig) **16** (2007) 218.

[34] S. M. Ikhdair and R. Sever, J. Phys. A: Math. Theor. **44** (2011) 345301.

[35] S. M. Ikhdair and R. Sever, Appl. Math. Comput. **216** (2010) 545; S. M. Ikhdair and R. Sever, Appl. Math. Comput. **216** (2010) 911.

[36] S. M. Ikhdair, C. Berkdemir and R. Sever, Appl. Math. Comput. **217** (2011) 9019; S. M. Ikhdair and R. Sever, Appl. Math. Comput. **218** (20) (2012) 10082.

[37] S. M. Ikhdair, Eur. J. Phys. A **39** (2009) 307.

[38] C. -S. Jia, T. Chen and L. -G. Cui, Phys. Lett. A 373 (2009) 1621.

[39] R. L. Greene and C. Aldrich, Phys. Rev. A **14** (1976) 2363.

[40] S. M. Ikhdair and R. Sever, J. Math. Chem. 42 (2007) 461.

[41] A. Arai, J. Math. Anal. Appl. **158** (1991) 63.

[42] H. Eğrifes, D. Demirhan and F. Büyükkılıç, Phys. Scr. **59** (1999) 90; **60** (1999) 195.

[43] I. S. Gradsteyn and I. M. Ryzhik, Tables of Integrals, Series and Products, Academic Press, New York, 1994.

[44] A. D. Alhaidari, Found. Phys. 40 (2010) 1088 (DOI: 10.1007/s10701-010-9431-5).

[45] Y. Xu, S. He and C. S. Jia, Phys. Scr. **81** (2010) 045001.

[46] J. N. Ginocchio, Phys. Rev. Lett. **78** (3) (1997) 436.

[47] J. Meng, K. Sugawara-Tanabe, S. Yamaji and A. Arima, Phys. Rev. C 59 (1999) 154.

[48] J. Meng, K. Sugawara-Tanabe, S. Yamaji, P. Ring and A. Arima, Phys. Rev. C 58 (1998) R628.

[49] S. M. Ikhdair and R. Sever, Cent. Eur. J. Phys. **8** (2010) 665.

[50] S. M. Ikhdair, J. Math. Phys. **52** (2011) 052303.

[51] S. M. Ikhdair, J. Math. Phys. **51** (2010) 023525.

[52] S. M. Ikhdair, Cent. Eur. J. Phys. **10** (2012) 361; M. Hamzawi and S. M. Ikhdair, Can. J. Phys. **90** (2012) 655.

[53] M. Hamzvai, S. M. Ikhdair and B. I. Ita, Phys. Scr. **85** (2012) 045009; M. Hamzavi, M. Eshghi and S. M. Ikhdair, J. Math. Phys. **53** (2012) 082101.

[54] P. Alberto, A. S. de Castro and M. Malheiro, Phys. Rev. C **75** (2007) 047303.

[55] S. M. Ikhdair and R. Sever, J. Math. Phys. **52** (2011) 122108.

**Table 1** A **sp**in symmetric **e**nergy levels of a Dirac particle subject to a complexified $PT$-symmetric Pöschl-Teller potential for various values $n$ and $\kappa$.

| $l$ | $n; \kappa$ | $(l, j = l+1/2)$ | $E_{n,\kappa}$ ($fm^{-1}$) |
|---|---|---|---|
| 1 | 0; -2,1 | $0p_{3/2}, 0p_{1/2}$ | 4.320628792 |
| 2 | 0; -3,2 | $0d_{5/2}, 0d_{3/2}$ | 4.451695423 |
| 3 | 0; -4,3 | $0f_{7/2}, 0f_{5/2}$ | 4.600717080 |
| 4 | 0; -5,4 | $0g_{9/2}, 0g_{7/2}$ | 4.751280043 |
| 1 | 1; -2,1 | $1p_{3/2}, 1p_{1/2}$ | 4.644277674 |
| 2 | 1; -3,2 | $1d_{5/2}, 1d_{3/2}$ | 4.739928403 |
| 3 | 1; -4,3 | $1f_{7/2}, 1f_{5/2}$ | 4.848061200 |
| 4 | 1; -5,4 | $1g_{9/2}, 1g_{7/2}$ | 4.955818811 |

**Table 2** The p-spin symmetric **e**nergy levels of a Dirac particle subject to a complexified $PT$-symmetric Pöschl-Teller potential for various values $n$ and $\kappa$.

| $\tilde{l}$ | $n, \kappa$ | $(l, j)$ | $E_{n,\kappa}$ ($fm^{-1}$) |
|---|---|---|---|
| 1 | 1, -1,2 | $1s_{1/2}, 0d_{3/2}$ | −5.170251165 |
| 2 | 1, -2,3 | $1p_{3/2}, 0f_{5/2}$ | −5.055448493 |
| 3 | 1, -3,4 | $1d_{5/2}, 0g_{7/2}$ | −4.943195896 |
| 4 | 1, -4,5 | $1f_{7/2}, 0h_{9/2}$ | −4.846340118 |
| 1 | 2, -1,2 | $2s_{1/2}, 1d_{3/2}$ | −5.000631769 |
| 2 | 2, -2,3 | $2p_{3/2}, 1f_{5/2}$ | −4.951890564 |
| 3 | 2, -3,4 | $2d_{5/2}, 1g_{7/2}$ | −4.915209098 |
| 4 | 2, -4,5 | $2f_{7/2}, 1h_{9/2}$ | −4.900619782 |

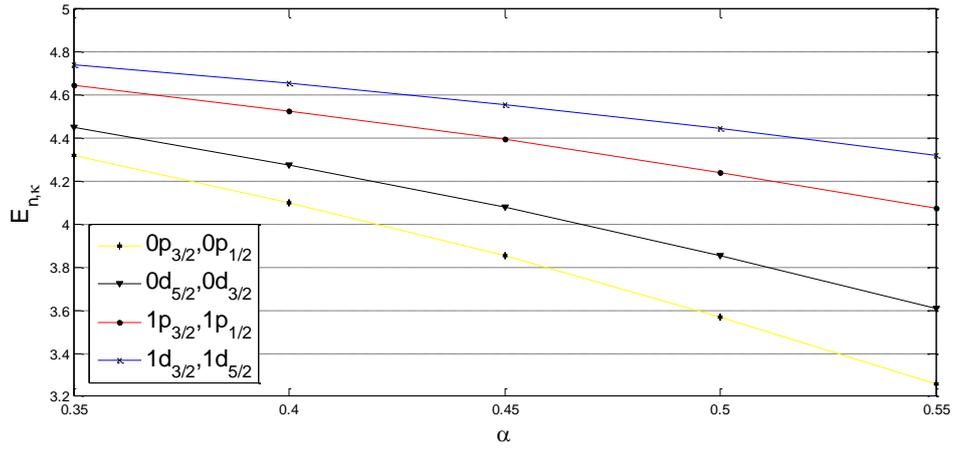

Fig. 1 The variation of the energy levels as a function $\alpha$ in view of spin symmetry with parameter values $A=8, B=2, M=5.0, C_s=0.35$.

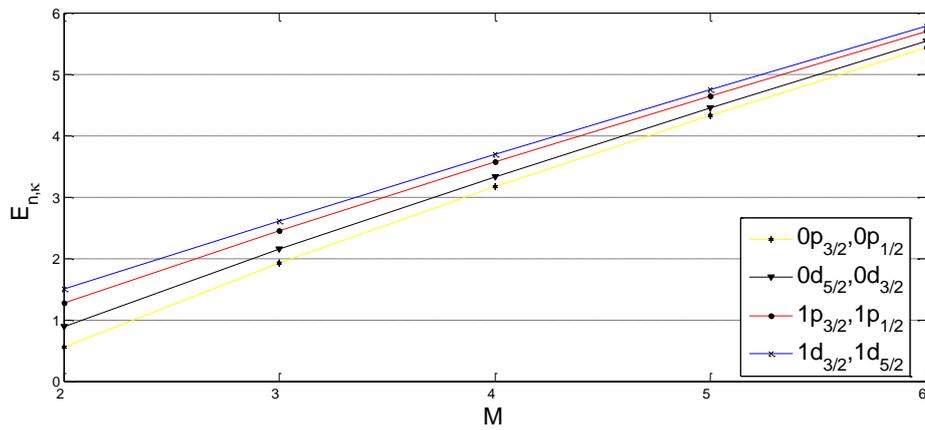

Fig. 2 The variation of the energy levels as a function $M$ in view of spin symmetry with parameter values $A=8, B=2, C_s=0.35, \alpha=0.35$.

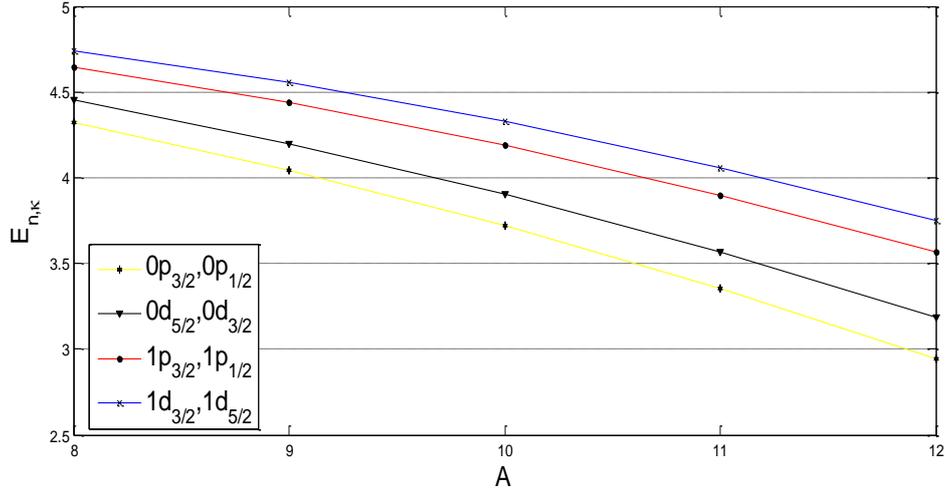

Fig. 3 The variation of the energy levels as a function $A$ in view of spin symmetry with parameter values $B=2, M=5.0, C_s=0.35, \alpha=0.35$.

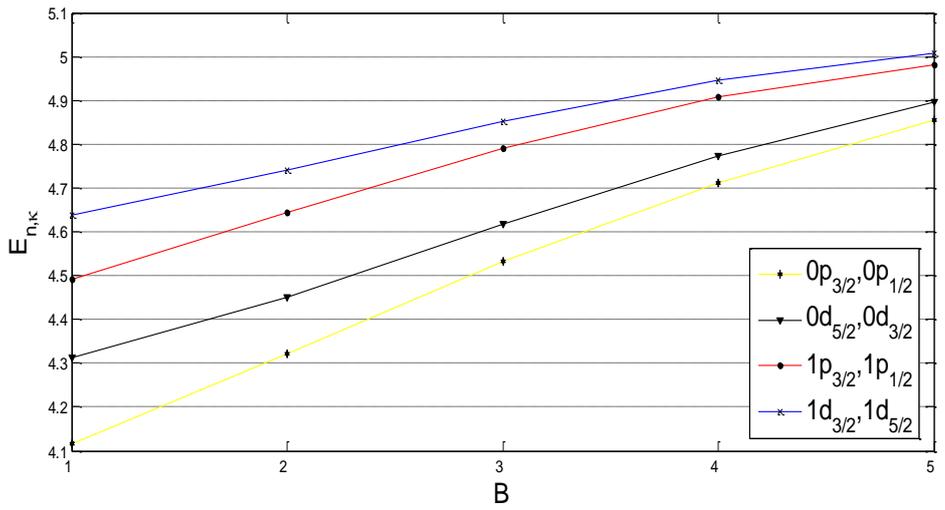

Fig. 4 The variation of the energy levels as a function $B$ in view of spin symmetry with parameter values $A=8, M=5.0, C_s=0.35, \alpha=0.35$.

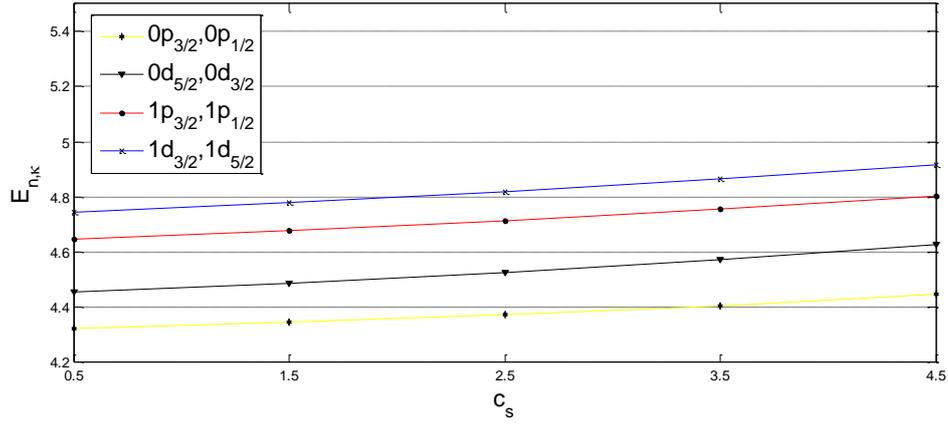

Fig. 5 The variation of the energy levels as a function $c_s$ in view of spin symmetry with parameter values $A=8, B=2, M=5.0, \alpha=0.35$.

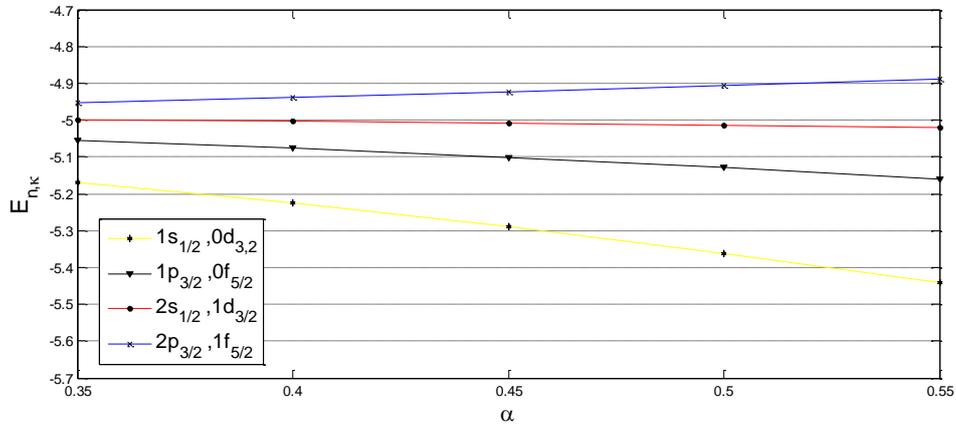

Fig. 6 The variation of the energy levels as a function $\alpha$ in the presence of p-spin symmetry taking . $A=8, B=2, M=5.0, C_{ps}=-15$.

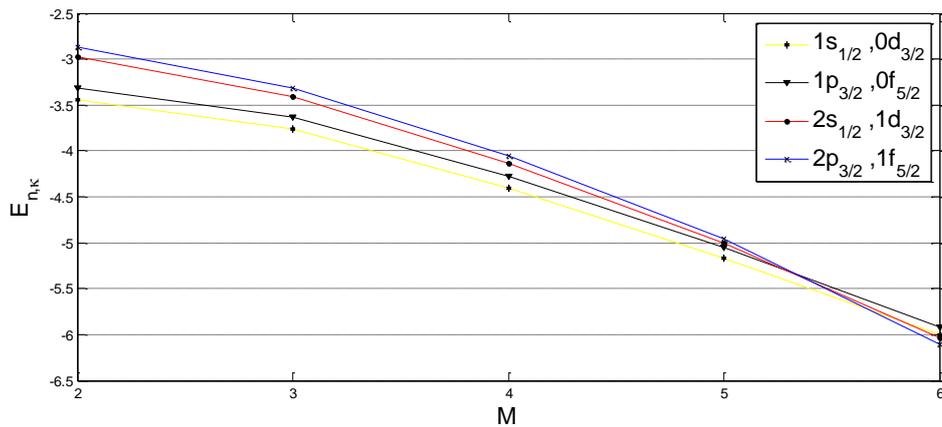

Fig. 7 The variation of the energy levels as a function $M$ in the presence of p-spin symmetry taking $A=8, B=2, C_{ps}=-15, \alpha=0.35$.

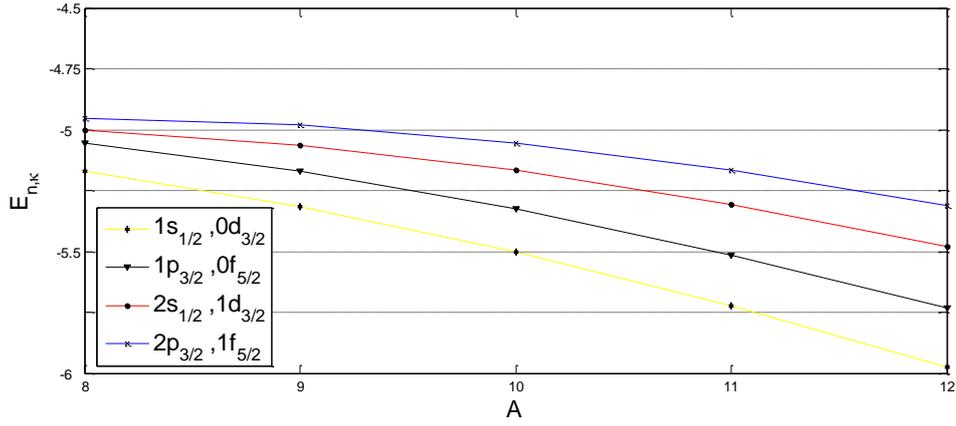

Fig. 8 The variation of the energy levels as a function $A$ in the presence of p-spin symmetry taking $B=2, M=5.0, C_{ps}=-15, \alpha=0.35$.

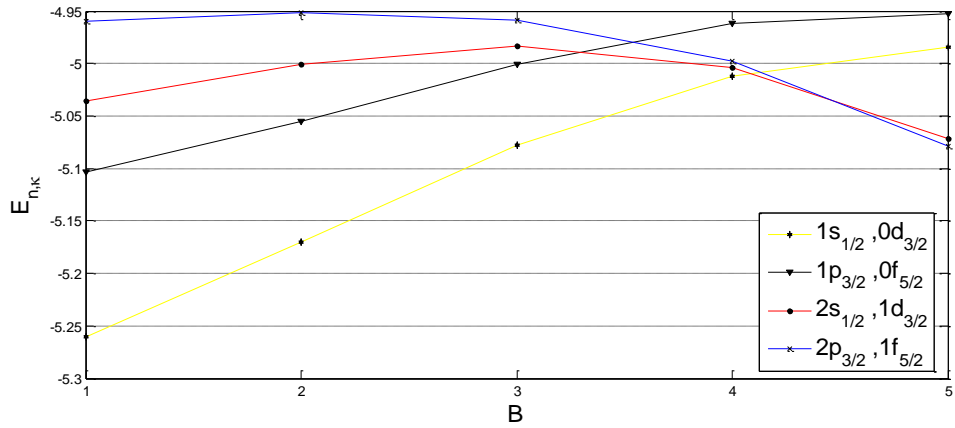

Fig. 9 The variation of the energy levels as a function $B$ in the presence of p-spin symmetry with parameter values $A=8, M=5.0, C_{ps}=0.35, \alpha=0.35$.

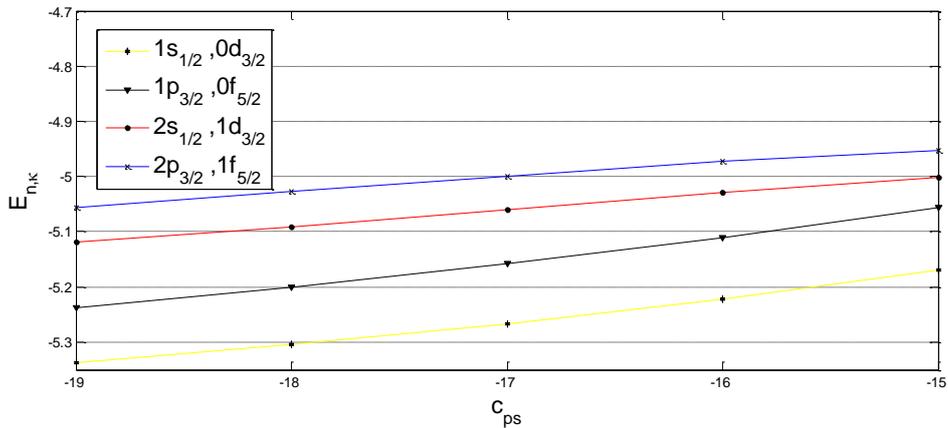

Fig. 10 The variation of the energy levels as a function $c_{ps}$ in the presence of p-spin symmetry with parameter values $A=8, B=2, M=5.0, \alpha=0.35$.